\documentstyle[12pt]{article}

\title{General relativity as an effective field theory:\\
The leading quantum corrections}

\author{John F. Donoghue\\ [5mm]
Department of Physics and Astronomy\\
University of Massachusetts, Amherst, MA  01002}
\date{ }
\begin{document}
\begin{titlepage}
\maketitle
\begin{abstract}

I describe the treatment of gravity as a quantum effective field theory.  This
allows a natural separation of the (known) low energy quantum effects
from the (unknown) high energy contributions.  Within this framework,
gravity is a well behaved quantum field theory at ordinary energies.
In studying the class of quantum corrections at low energy, the
dominant effects at large distance can be isolated, as these are due to the
propagation of the massless particles (including gravitons) of the theory
and are manifested in the nonlocal/nonanalytic contributions to vertex
functions and propagators. These leading quantum corrections are
parameter-free and represent necessary consequences of quantum gravity.
The methodology is illustrated by a calculation of the leading quantum
corrections to the gravitational interaction of two heavy masses.

\end{abstract}
{\vfill UMHEP-408}
{\vfill gr-qc/9405057 }
\end{titlepage}

\section{Introduction}

We are used to the situation where our theories are only assumed to be
provisional.
They have been tested and found to be valid over a limited range of energies
and
distances.  However, we do not know that they hold in more extremes
situations.
There are many examples of theories which have been superseded by new
theories
at higher energies, and we expect this process to continue.  It is interesting
to look
at the incompatibility of general relativity and quantum mechanics in this
light.  It
would not be surprising if there are new ingredients at high energy in order
to have
a satisfactory theory of quantum gravity.  However, are there any conflicts
between
gravity and quantum mechanics at the energy scales that are presently
accessible?  If
there are, it would be a major concern because it would mean our present
theories
are wrong in ways which cannot be blamed on new physics at high energy.

There is an apparent technical obstacle to addressing the compatibility of
quantum
mechanics and gravity at present energies, i.e., the non renormalizability of
quantum gravity.  Quantum fluctuations involve all energy scales, not just
the
energy of the external particles.  Perhaps our lack of knowledge of the true
high
energy theory will prevent us from calculating quantum effects at low
energy.  In
the class of renormalizable field theories, low energy physics is shielded
from this
problem because the high energy effects occur only in the shifting of a small
number of parameters.  When these parameters are measured
experimentally, and
results expressed in terms the measured values, all evidence of high energy
scales
disappears or is highly suppressed.[1]  However in some
nonrenormalizable
theories,
the influence of high
energy remains.  For example in the old Fermi theory of weak interactions,
the ratio
of the neutron decay rate to that of the muon has a contribution which
diverges
logarithmically at one loop.  It is not the divergence itself which is the
problem, as
the ratio becomes finite in the Standard Model (with a residual effect of
order
$\alpha \, ln \, M^2_Z$).  More bothersome is the sensitivity to the high
energy
theory--the low energy ratio depends on whether the scale of the new
physics is
$M_Z$ or $10^{14} GeV$.

However quantum predictions can be made in non-renormalizable theories.
The
techniques are those of effective field theory, which has been assuming
increasing
importance as a calculational methodology.  The calculations are organized
in a
systematic expansion in the energy.  Effects of the high energy theory again
appear
in the form of shifts in parameters which however are determined from
experiment.
To any given order in the energy expansion there are only a finite number of
parameters, which can then be used in making predictions.  Using the
techniques of
effective field theory, it is easy to separate the effects due to low energy
physics
from that of the (unknown) high energy theory.  Indeed, even the phrasing
of the
question raised
in
the opening paragraph is a by-product of the way of thinking about effective
field
theory.

General relativity fits naturally into the framework of effective field
theory. The
gravitational interactions are proportional to the energy, and are easily
organized
into an energy expansion.  The theory has been quantized on smooth
enough
background metrics.[2,3,4]  We will explore quantum gravity as an
effective field
theory
and find no obstacle to its successful implementation.

In the course of our study we will find a class of quantum predictions which
are
parameter free (other than Newton's constant G) and which dominates over
other
quantum predictions in the low energy limit.  These 'leading quantum
corrections'
are the first modifications due to quantum mechanics, in powers of the
energy or
inverse factors of the distance.  Because they are independent of the
eventual high
energy theory of gravity, depending only on the massless degrees of
freedom and
their low energy couplings, these are true predictions of quantum general
relativity.

The plan of the paper is as follows.  In Section 2, we briefly review general
relativity and its quantization.  Section 3 is devoted to the treatment of
general
relativity as an effective field theory, while the leading quantum corrections
are
described in more detail in Section 4.  We give more details of a previously
published example,[5] that of the gravitational interaction around flat space,
in
Section
5.  Some speculative comments on the {
\em extreme} low energy limit, where the
wavelength is of order the size/life time of the universe are given in Section
6 with
concluding comments in Section 7.  An appendix gives some of the non
analytic
terms needed for the leading quantum corrections arising in loop integrals.

\section{General Relativity and its Quantization}

In this paper the metric convention is such that flat space is represented by
$\eta_{\mu \nu} = diag (1, -1, -1, -1)$.[6,7]  The Einstein action is

\begin{equation}
S_{grav} = \int d^4x \sqrt{-g} \left[ {2 \over \kappa{^2}} R \right]
\end{equation}

\noindent where $\kappa^2 = 32 \pi G, g=det g_{\mu \nu}, g_{\mu \nu}$
is the
metric
tensor and $R=g^{\mu \nu} R_{\mu \nu}$.

\begin{eqnarray}
R_{\mu \nu} & = & \partial_{\nu} \Gamma^{\lambda}_{\mu \lambda} -
\partial_{\lambda} \Gamma^{\lambda}_{\mu \nu} +
\Gamma^{\sigma}_{\mu
\lambda} \Gamma^{\lambda}_{\nu \sigma} - \Gamma^{\sigma}_{\mu
\nu}
\Gamma^{\lambda}_{\lambda \sigma} \nonumber \\
\Gamma^{\lambda}_{\alpha \beta} & = & {g^{\lambda \sigma} \over 2}
\left(
\partial_{\alpha} g_{\beta \sigma} + \partial_{\beta} g_{\alpha \sigma} -
\partial_{\sigma} g_{\alpha \beta} \right)
\end{eqnarray}

\noindent Heavy spinless matter fields interact with the gravitational field as
described by the action

\begin{equation}
S_{matter} = \int d^4x \sqrt{-g} \left[ {1 \over 2} g^{\mu \nu}
\partial_{\mu}
\phi \partial_{\nu} \phi - {1 \over 2} m^2 \phi^2 \right]
\end{equation}

The quantum fluctuations in the gravitational field may be expanded about a
smooth background metric $\bar{g}_{\mu \nu}$, with one possible choice
being

\begin{eqnarray}
g_{\mu \nu} & = & \bar{g}_{\mu \nu} + \kappa h_{\mu \nu} \nonumber \\
g^{\mu \nu} & = & \bar{g}^{\mu \nu} - \kappa h^{\mu \nu} + \kappa{^2}
h^{\mu}_{\lambda}
h^{\lambda \nu} + \ldots
\end{eqnarray}

\noindent Indices here and in subsequent formulas are raised and lowered
with the
background metric.  In order to quantize the field $h_{\mu \nu}$, we need
to fix
the gauge.
Following 't~Hooft and Veltman[3] this entails a gauge fixing term

\begin{equation}
{\cal L}_{gf} = \sqrt{-\bar{g}} \left\{\left( h^{~~; \nu}_{\mu \nu} - {1 \over
2} h_{; \mu} \right) \left( h^{\mu \lambda}_{~~; \lambda} - h^{; \mu} \right)
\right\}
\end{equation}

\noindent with $h \equiv h^{\lambda}_{\lambda}$, and with the semicolon
denoting covariant differentiation on the background metric.
The ghost Lagrangian is

\begin{equation}
{\cal L}_{ghost} = \sqrt{-\bar{g}} \eta^{\ast \mu} \left\{ \eta^{~~~;
\lambda}_{\mu ;
\lambda} - \bar{R}_{\mu \nu}
\eta^{\nu} \right\}
\end{equation}

\noindent for the complex Faddeev-Popov ghost field $\eta_\mu$.

The expansion of the Einstein action takes the form [3,8]

\begin{equation}
S_{grav} = \int d^4x \sqrt{-\bar{g}} \left[ {2 \bar{R} \over
\kappa^2} +
{\cal
L}^{(1)}_g + {\cal L}^{(2)}_g + \ldots \right]
\end{equation}

\noindent where

\begin{eqnarray}
{\cal L}^{(1)}_g & = & {h_{\mu \nu} \over \kappa} \left[ \bar{g}^{\mu
\nu}
\bar{R} -
2 \bar{R}^{\mu \nu} \right] \nonumber \\
{\cal L}^{(2)}_g & = &  {1 \over 2} h_{\mu \nu ; \alpha} h^{\mu \nu ;
\alpha} -
{1 \over 2} h_{; \alpha} h^{; \alpha} + h_{; \alpha} h^{\alpha \beta}_{~~;
\beta} -
h_{\mu \beta ; \alpha} h^{\mu \alpha ; \beta} \nonumber \\
& & + \bar{R} \left( {1 \over 4} h^{2} - {1 \over 2} h_{\mu \nu} h^{\mu
\nu}
\right) + \left( 2 h^{\lambda}_{\mu} h_{\nu \lambda} - h h_{\mu \nu}
\right)
\bar{R}^{\mu \nu}
\end{eqnarray}

\noindent   A similar
expansion of
the matter action yields

\begin{equation}
S_{matter} = \int d^4x \sqrt{-\bar{g}} \left\{ {\cal L}^0_m + {\cal L}^{(1)}_m
+ {\cal
L}^{(2)}_m + \ldots \right\}
\end{equation}

\noindent with

\begin{eqnarray}
{\cal L}^{(0)}_m & = & {1 \over 2} \left( \partial_{\mu} \phi
\partial^{\mu} \phi
- m^2 \phi^2 \right) \nonumber \\
{\cal L}^{(1)}_m & = & - {\kappa \over 2} h_{\mu \nu} T^{\mu \nu}
\nonumber
\\
T_{\mu \nu} & \equiv &  \partial_{\mu} \phi \partial_{\nu} \phi - {1 \over
2}
\bar{g}_{\mu \nu} \left( \partial_{\lambda} \phi \partial^{\lambda} \phi -
m^2
\phi^2 \right) \nonumber \\
{\cal L}^{(2)}_m & = &  \kappa{^2} \left( {1 \over 2} h^{\mu \nu}
h^{\nu}_{\lambda}
- {1 \over 4} h h^{\mu \nu} \right) \partial_{\mu} \phi \partial_{\nu} \phi
\nonumber \\
& & - {\kappa{^2} \over 8} \left( h^{\lambda \sigma} h_{\lambda \sigma}
- {1 \over 2} h h
\right)
\left[
\partial_{\mu} \phi \partial^{\mu} \phi - m^2 \phi^2 \right]
\end{eqnarray}

\noindent If the background metric satisfies Einstein's equation

\begin{equation}
\bar{R}^{\mu \nu} - {1 \over 2} \bar{g}^{\mu \nu} \bar{R} = -{\kappa^2
\over 4}
T^{\mu \nu}
\end{equation}

\noindent the linear terms in $h_{\mu \nu}, {\cal L}^{(1)}_g + {\cal
L}^{(1)}_m$, will vanish.

In calculating quantum corrections at one loop, we need to consider the
Lagrangian
to quadratic
order

\begin{eqnarray}
S_0 & = & \int d^4 x \sqrt{-g} \left\{{2 \bar{R} \over \kappa^2} + {\cal
L}^{(0)}_m \right\}
\nonumber \\
S_{quad} & = & \int d^4 x \sqrt{-g} \left\{ {\cal L}^{(2)}_g + {\cal
L}_{gf} +
{\cal L}_{ghost}
+ {\cal L}^{(2)}_m \right\}
\end{eqnarray}

\noindent The integration over the gravitational degrees of freedom

\begin{equation}
W [\phi] = e^{iz[\phi]} = \int [d h_{\mu \nu}] [d \eta_{\mu}] e^{i(S_0 +
S_{quad})}
\end{equation}

\noindent yields a functional which in general is non-local and also
divergent.  The
identification of
the quantum degrees of freedom and the definition of a quantum theory is
no more
difficult than the
quantization of Yang Mills theory, at least for small quantum fluctuations.
The
difficulties arise in
applying this result.  Because of the dimensionful coupling $\kappa$ and
the
nonlinear nature of
the theory, divergencies appear in places which cannot be absorbed into the
basic
parameters
introduced this far.  Since the coupling grows with energy, the theory is
strongly
coupled at very
high energy, $E > M_{Planck}$, and badly behaved in perturbation theory.
We
also do not have
known techniques for dealing with large fluctuations in the metric, which
may in
principle be
topology-changing in nature.  However, the low energy fluctuations are
very
weakly coupled, with
a typical strength $\kappa^2 q^2 \sim 10^{-40} [10^{-70}]$ for $q^2 \sim
1 / (1
fm)^2 \left[ q^2
\sim 1 / (1 m)^2 \right]$.  Since small quantum fluctuations at ordinary
energies
behave normally
in perturbation theory, it is natural to try to separate these quantum
corrections from
the
problematic (and most likely incorrect) high energy fluctuations.  Effective
field
theory is the tool
to accomplish this separation.

\section{Gravity as an Effective Field Theory}

Effective field theory techniques[9,10] have become common in particle
physics.
The method is
not a
change in the rules of quantum mechanics, but is rather a procedure
which organizes the
calculation and
separates out the effects of high energy from the known
quantum effects at low
energy.  General
relativity is a field theory which naturally lends itself to such a treatment.
Perhaps
the known manifestation of a effective field theory which is closest in style
to
gravity is chiral perturbation theory,[10] representing the low energy limit
of QCD.
Both are nonlinear, nonrenormalizable theories with a dimensionful
coupling
constant.  If the pion mass were taken to zero, as can be easily achieved
theoretically, long distance effects similar to those from graviton loops
would be
found.  In addition we have had the benefit of detailed calculations and
experimental
verification in the chiral case, so that the workings of effective field theory
are
transparent.  In this section, I transcribe the known properties of effective
field
theory to the gravitational system.

The guiding principle underlying general relativity is that of gauge
symmetry, i.e.,
the local invariance under coordinate transformations.  This forces the
introduction
of geometry, and requires us to define the action of the theory using
quantities
invariant under the general coordinate transformations.  However this is not
sufficient to completely define the theory, as many quantities are invariant.
For
example, each term in the action

\begin{equation}
S_{grav} = \int d^4x \sqrt{-g} \left\{ \Lambda + {2 \over \kappa^2} R + c_1
R^2 + c_2
R_{\mu \nu} R^{\mu \nu} + \dots \right\} ,
\end{equation}

\noindent (where $\Lambda, \kappa^2, c_1, c_2$ are constants and the ellipses
denote higher powers of $R, R_{\mu \nu}$ and $R_{\mu \nu \alpha
\beta}$), are
separately invariant under general coordinate transformations.  Other
physics
principles must enter in order to simplify the action.  For example, the
constant
$\Lambda$ is proportional to the cosmological constant $(\lambda = - 8 \pi
G
\Lambda)$, which experiment tells us is very small.[11]  We therefore set
(the
renormalized value of) $\Lambda = 0$ for the rest of this paper.
Experiment tells
us very little about the dimensionless constants $c_1, c_2$, bounding $c_1,
c_2
\leq
10^{74}$,[12]
and coefficients of yet higher powers of R have essentially no
experimental constraints.  Einstein's theory can be obtained by setting $c_1,
c_2
= 0$ as well as forbidding all higher powers.  However it is very unlikely
that in
fact $c_1, c_2 = 0$.  For example, quantum corrections involving the
known
elementary particles (whether or not gravity itself is quantized) will generate
corrections to $c_1, c_2$ etc.  Unless an infinite number of "accidents"
occur
 at least some of the higher order coefficients will be nonzero.

In practice there is no known reason to require that $c_1, c_2$ vanish
completely.  We can instead view the gravitational action as being organized
in an
energy expansion, and then reasonable values of $c_1, c_2$ do not
influence
physics at low energies.  In order to set up the energy expansion, we note
that the
connection $\Gamma^{\lambda}_{\alpha \beta}$ is first order in derivatives
and the
curvature
is second order.  When matrix elements are taken, derivitives turn into
factors of the
energy or momentum $i \partial_{\mu} \sim p_{\mu}$, so that the curvature
is
said to be of order $p^2$.  Terms in the action involving two powers of the
curvature are of order $p^4$.  The graviton energy can be arbitrarily small
and at
low enough energies terms of order $p^4$ are small compared to those of
order
$p^2$.  The higher order Lagrangians will have little effect at low energies
compared to the Einstein term, $R$.  This is why the experimental bounds
on
$c_1, c_2$ are so poor; reasonable values of $c_1, c_2$ give modifications
which are unmeasurably feeble.  In a pure gravity theory the expansion
scale might
be expected to be the Planck mass $M^2_{pl} \simeq G^{-1}_N \approx
(1.2
\times 10^{19} GeV)^2$.  For example, if we just consider the $R$ and
$R^2$
terms in the Lagrangian, Einstein's equation is modified to

\begin{eqnarray}
\left( 1 + 32 \pi c_1 G_N R \right) \left(R_{\mu \nu} - {1 \over 2} g^{\mu
\nu} R
\right) & & \nonumber \\
 + 32 \pi c_1 G_N \left[ R_{; \lambda}^{~~\lambda} - R_{; \mu \nu} + {1
\over 4}
g_{\mu \nu}  R^2 \right] &=& - 8 \pi G T^{\mu \nu}
\end{eqnarray}

\noindent Unless $c_1R$ or $c_1 R_{; \lambda}^{~~\lambda} / R \geq
10^{70}
m^{-2}$ the
influence of the $c_1
R^2$ terms is negligible.

In the literature[13] there are discussions of problems with $R + R^2$
theories.
These include negative metric states, unitarity violation, an inflationary
solution,
and an instability of flat space.  However J. Simon[14] has shown that
these
problems
do not appear when the theory is treated as an effective field theory.
Essentially,
the problems arise from treating the $R + R^2$ description (without any
higher
terms) as a fundamental theory at high energy when the curvature is of order
the
Planck mass squared.  Then the $R^2$ contribution is comparable to that of
$R$.
Of course then yet higher powers of $R$ would also be comparable to the
$R^2$
and $R$ terms, so that in this region we would not be able to say anything
about
the full $R+R^2 + R^3 + R^4 + \dots$ expansion.  In the low energy
region the
effect of $R^2$ is just a small correction to the behavior of the pure Einstein
theory
and no bad behavior is introduced.

The most general gravitational action will have an infinite number of
parameters
such as $\kappa^2, c_1, c_2$.  At the lowest energy, only $\kappa^2$ is
important.
However we can imagine in principle that we could do experiments with
such high
precision that we could also see the first corrections and measure $c_1,
c_2$.  If
we knew the ultimate correct theory of gravity, we might be able to predict
$\kappa,
c_1, c_2$.  With our incomplete knowledge at low energy, we must treat
them
as free parameters.  Quantum effects from both high energy and low energy
particles have the potential to produce shifts in $\kappa, c_1, c_2$ and it is
the final
(renormalized) value which experiments would determine.

It is crucial to differentiate the quantum effects of heavy particles from
those of
particles which are massless.  The issue is one of scale.  Virtual heavy
particles
cannot propagate long distances at low energies; the uncertainty principle
gives
them a range $\Delta r \sim 1/M_H$.  On distance scales much larger than
this, their effects will look local, as if they were described by a local
Lagrangian.
Even the slight nonlocality can be accounted for by Taylor expanding the
interactions about a point.  In a simple example, a particle propagator can be
Taylor
expanded

\begin{equation}
{1 \over q^2 - M^2_H} = {-1 \over M^2_H} - {q^2 \over M^4_H} - {q^4
\over
M^6_H} +
\ldots
\end{equation}

In the coordinate space propagator obtained by a Fourier transform, the
constant
$1/M^2_H$ generates a delta function, hence a local interaction and the
factors of
$q^2$ are replaced by derivatives in a local Lagrangian.  The quantum
effects of
virtual heavy particles then appear as shifts in the coefficients of most
general
possible local Lagrangian.

On the other hand, the quantum effects of massless particles cannot all be
accounted
for in this way.  Some portions of their quantum corrections, for example
the
results of high energy propagation in loop integrals, do shift the parameters
in the
Lagrangian and are local in that respect.  However the low energy
manifestations of
massless particles are not all local, as the particles can propagate for long
distances.
A simple example is again the propagator, now $1/q^2$, which cannot be
Taylor
expanded about $q^2 = 0$.  The low energy particles (massless ones or
ones
whose mass is comparable to or less than the external energy scale) cannot
be
integrated out of the theory but must be included explicitly in the quantum
calculations.

Our direct experience in physics covers distances from $10^{-17}m$ to
cosmic
distance scales.
Although gravity is not well tested over all of those scales, we would like
to believe that both general relativity and quantum mechanics are valid in
that range,
with likely modifications coming in at $\sim 10^{-39}m \sim 1/M_{pl}$.
For
reasons discussed more fully in Section 6, I would like to imagine
quantizing the
theory in a very large, but not infinite, box.  Roughly speaking, this is to
avoid
asking questions about wavelengths of order the size of the Universe, i.e.,
reaching
back to the Big Bang singularity.  However, the volume is taken large
enough that
we may ignore edge effects.  We assume that any particles which enter this
quantization volume (e.g., the remnants of the Big Bang) are either known
or
irrelevant.  The curvature is assumed to be small and smooth throughout
this space-
time volume.  This situation then represents the limits of our "known"
confidence in
both general relativity and quantum mechanics, and we would like to
construct a gravitational effective field theory (GEFT) in this region.

The dynamical information about a theory can be obtained from a path
integral.  The
results of the true theory of gravity will be contained in a generating
functional

\begin{equation}
W [J ] = e^{Z[J]} = \int \left[ d \phi \right] \left[d (gravity) \right] e^{i
S_{true}
(\phi, (gravity), J)}
\end{equation}

\noindent where (gravity) represents the fields of a full gravity theory,
$\phi$
represents matter fields and $J$ can be a set of source fields added to the
Lagrangian (i.e., $\Delta {\cal L} = -J \phi$) in order to allow us to probe
the
generating functional.  The gravitational effective field theory is defined to
have the
same result

\begin{equation}
W[J] = e^{Z[J]} = \int [d \phi] \left[ d \, h_{\mu \nu} \right] e^{i S_{eff}
(\phi,
\bar{g}, h, J)}
\end{equation}

\noindent Here $S_{eff}$ is constructed as the most general possible
Lagrangian
containing $g, \phi$ and $J$ consistent with general covariance.  It contains
an
infinite number of free parameters, such as $\kappa, c_1, c_2$ described
above.
The effects of the high energy part of the true theory are accounted for in
these
constants.  However, as discussed above, the low energy degrees of
freedom must
be accounted for explicitly, hence their inclusion in the path integral.
Since we are
only interested
in the small fluctuation and low energy configurations of $h_{\mu \nu}$, we
do not
need to address
issues of the functional measure for large values of $h_{\mu \nu}$.  Any
measure
and
regularization scheme which does not violate general covariance may be
used.
Because the
coupling of the low energy fluctuations is so very weak, the path integral
has a well
behaved
perturbative expansion.  We have
implicitly assumed that gravitons are the only low energy particles which are
remnants of the full gravity theory.  If any other massless particles result,
they
would also need to be included.

The most general effective Lagrangian will contain both gravitational and
matter
terms and will
written in a derivative expansion

\begin{eqnarray}
S_{eff} &=& \int d^4 x \sqrt{-g} {\cal L}_{eff} \nonumber \\
{\cal L}_{eff} &=& {\cal L}_{grav} + {\cal L}_{matter} \nonumber \\
{\cal L}_{grav} &=& {\cal L}_{g0} + {\cal L}_{g2} + {\cal L}_{g4} +
\ldots
\nonumber \\
{\cal L}_{matter} &=& {\cal L}_{m0} + {\cal L}_{m2} + \ldots
\end{eqnarray}

\noindent The general gravitational component has already been written
down

\begin{eqnarray}
{\cal L}_{g0} &=& \Lambda \nonumber \\
{\cal L}_{g2} &=& {2 \over \kappa^2} R \nonumber \\
{\cal L}_{g4} &=& c_1 R^2 + c_2 R_{\mu \nu} R^{\mu \nu}
\end{eqnarray}

\noindent The first two terms in the general matter Lagrangian for a massive
field
are

\begin{eqnarray}
{\cal L}_{m0} &=& {1 \over 2} \left[ g^{\mu \nu} \partial_{\mu} \phi
\partial_{\nu} \phi - m^2
\phi^2 \right] \nonumber \\
{\cal L}_{m2} &=& d_1 R^{\mu \nu} \partial_{\mu} \phi \partial_{\nu}
\phi + R
\left( d_2
\partial_{\mu} \phi \partial^{\mu} \phi + d_3 m^2 \phi^2 \right)
\end{eqnarray}

\noindent Note that derivatives acting on a massive matter field $\phi$ are
not small
quantities--the
ordering in the derivative expansion only counts derivatives which act on
the light
fields, which in
this case is only the gravitons.  In contrast, if the matter field were also
massless,
the ordering in
energy would be different

\begin{eqnarray}
\bar{\cal L}_{m0} &=& 0 \nonumber \\
\bar{\cal L}_{m2} &=& {1 \over 2} g^{\mu \nu} \partial_{\mu} \phi
\partial_{\nu} \phi +
\bar{d}_3 R
\phi^2 \nonumber \\
\bar{\cal L}_{m4} &=& \bar{d}_1 R^{\mu \nu} \partial_{\mu} \phi
\partial_{\nu}
\phi +
\bar{d}_2 R
\partial_{\mu} \phi \partial^{\mu} \phi + \left( \bar{d}_4 R^2 + \bar{d}_5
R_{\mu
\nu} R^{\mu
\nu} \right) \phi^2
\end{eqnarray}

\noindent where the overbar is meant to indicate the parameters in the
massless
theory.  For
$\bar{d}_3 = {-1 \over 12}$, we have the "improved" action of Callen,
Coleman
and Jackiw[15];
however any value of $\bar{d}_3$ is consistent with general covariance and
the
energy expansion.
Note that since we are only working to ${\cal O} (p^4)$, we may use the
lowest
order equations
of motion to simplify ${\cal L}_{m2}, \bar{{\cal L}}_{m4}$.

Let us now discuss the expected size of the parameters in the effective
action.
Those parameters
which are accessible to realistic measurements $(\Lambda, \kappa, m)$ have
been
labeled by their
conventional names.  From the standpoint of the energy expansion it is of
course a
great shock that
the renormalized value of the cosmological constant is so small $(\lambda =
-8 \pi G
\Lambda)$.
The cosmological bound is $\mid \Lambda \mid = 10^{-46} GeV^4$,[11]
where as
the standard
expectation would be a value 40 to 60 orders of magnitude larger.  Effective
field
theory has
nothing special to say about this puzzle.  However it does indicate that at
ordinary
scales,
$\Lambda$ is unimportant and that the energy expansion for gravity starts at
two
derivatives with
${\cal L}_{g2}$.  The constants $c_1, c_2$ are dimensionless.  They
determine
the scale of the
energy expansion of pure gravity which in general is

\begin{equation}
1 + \kappa^2 q^2 c_i \approx 1 + {q^2 \over \Lambda^2_{grav}}
\end{equation}

\noindent We of course have no direct experience with this scale, but the
expectation that
$\Lambda_{grav}$ is of
order of the Planck mass would correspond to $c_1, c_2 \approx {\cal
O}(1)$.
The
phenomenological bound [12],  $c_1, c_2 \leq 10^{74}$, is of course
nowhere
close to being able
to
probe this possibility.

For the constants in the matter Lagrangian, $d_i$, the expectations are a bit
more
complicated, as
we must distinguish between point particles which have only gravitational
interactions and those
which have other interactions and/or a substructure.  The constants $d_i$
have
dimension
$1/(mass)^2$, and we will see by explicit calculation in Section V that they
contribute to the form
factors in the energy-momentum vertex for the $\phi$ particle, being
equivalent to
the charge radius
in the well known electromagnetic form factor.  Loop diagrams involving
gravitons
shift $d_i$ by
terms of order $\kappa^2 \approx 1 / M^2_{p \ell}$.  In the absence of
interactions other than
gravity,
it  is consistent to have $d_i \approx {\cal O} \left( 1/M^2_{p \ell} \right)$.
However, for
particles  that have other interactions, the energy and momentum will be
spread out
due to quantum
fluctuations and a gravitational charge radius will result.  The expectation in
this
case is

\begin{equation}
d_1, d_2, d_3 \approx {\cal O} \left( \langle r^2 \rangle_{grav} \right)
\end{equation}

\noindent For composite particles, such as the proton, this will be of size of
the
physical radius,
$\langle r^2 \rangle^{proton}_{grav} \approx 1 fm^2$.  For interacting
point
particles, such as the
electron, this would be of order the scale of quantum correction $d_1, d_2,
d_3
\approx \alpha /
m^2_e$.

The gravitational effective field theory is a full quantum theory and loop
diagrams
are required in order to satisfy general principles, such as unitarity.  The
finite
infrared part of loop diagrams will be discussed more fully in the next
section.
Also obtained in the usual calculation of many loop diagrams are ultraviolet
divergences.  These arise in a region where the low energy effective theory
may not
be reliable, and hence the divergences may not be of deep significance.
However
as a technical matter they must be dealt with without influencing low energy
predictions.  The method is known from explicit calculations of the
divergences in
gravity [3,4,16,17], and from general effective field theory practice.  The
divergences are
consistent with the underlying general covariance, and must occur in forms
similar
to terms in the most general possible effective Lagrangian.  They can then
be
absorbed into renormalized values for the unknown coefficients which
appear in
this general Lagrangian.  Moreover, it can be shown that for loops
involving low
order terms in the energy expansion, that the renormalization involves the
coefficients which appear at higher order.

In a background field method, one can compactly summarize the one loop
quantum
corrections.
The classical background field $\bar{g} (x)$ is determined by the matter
fields and
sources by the
equations of motion

\begin{equation}
\bar{R}_{\mu \nu} - {1 \over 2} \bar{g}_{\mu \nu} \bar{R} = 16 \pi G
{\delta
S_{eff,matter}
\over
\delta \bar{g}^{\mu \nu}} \mid_{h_{\alpha \beta} = 0}
\end{equation}

\noindent The effective action is thus rendered into quadratic form in
$h_{\mu
\nu}$,

\begin{equation}
S_{eff} = \int d^4_x \sqrt{\bar{g}} \left\{ \bar{{\cal L}} (\bar{g}) - {1
\over 2}
h_{\alpha \beta}
D^{\alpha \beta \gamma \delta} h_{\gamma \delta} + \ldots \right\}
\end{equation}

\noindent where

\begin{eqnarray}
D^{\alpha \beta \gamma \delta} &=& I^{\alpha \beta, \mu \nu} d_{\lambda}
d^{\lambda} I_{\mu
\nu,}^{~~\gamma \delta} - {1 \over 2} \bar{g}^{\alpha \beta} d_{\lambda}
d^{\lambda}
\bar{g}^{\gamma \delta} + \bar{g}^{\alpha \beta} d^{\gamma} d^{\delta}
+
d^{\alpha} d^{\beta}
\bar{g}^{\gamma \delta} \nonumber \\
& & -2I^{\alpha \beta, \mu \nu} d_{\sigma} d_{\lambda}
I_{\mu}^{~\sigma,
\gamma \delta} +
\bar{R} \left( I^{\alpha \beta, \gamma \delta} - {1 \over 2} g^{\alpha \beta}
g^{\gamma \delta}
\right) \nonumber \\
& & + \left( \bar{g}^{\alpha \beta} \bar{R}^{\gamma \delta} +
\bar{R}^{\alpha
\beta}
\bar{g}^{\gamma \delta} \right) - 4 I^{\alpha \beta, \lambda \mu}
\bar{R}_{\mu
\nu}
I_{\lambda}^{~\nu, \gamma \delta}
\end{eqnarray}

\noindent with $d_{\mu}$ being a covariant derivative and

\begin{equation}
I^{\alpha \beta, \gamma \delta} = {1 \over 2} \left( \bar{g}^{\alpha
\gamma}
\bar{g}^{\beta
\delta} + \bar{g}^{\alpha \delta} \bar{g}^{\beta \gamma} \right)  .
\end{equation}

\noindent Formally integrating over $h_{\mu \nu}$ one finds

\begin{equation}
Z [ \phi, J ] = Tr ln D
\end{equation}

While some of the finite portions of Z are difficult to extract (see the next
section),
since $ln D$ is
in general a nonlocal functional, the divergences are local and are readily
calculated.  One loop
divergences are known for gravity coupled to matter fields and the two loop
result has been found for pure gravity.  At one loop, the divergences due to
gravitons have the form [3]

\begin{equation}
\Delta {\cal L}^{(1)}_0 = {1 \over 8 \pi^2} {1 \over \epsilon} \left\{{1
\over 120}
R^2 + {7 \over 20} R_{\mu \nu} R^{\mu \nu} \right\}
\end{equation}

\noindent where $\epsilon = 4-d$ within dimensional regularization.  This
produces
the following
minimal subtraction renormalization of the gravitational parameters

\begin{eqnarray}
c^{(r)}_1 = c_1 + {1 \over 960 \pi^2 \epsilon} \\ \nonumber
c^{(r)}_2 = c_2 + {7 \over 160 \pi^2 \epsilon}
\end{eqnarray}

\noindent At two loops, the divergence of pure gravity is[16]

\begin{equation}
\Delta {\cal L}^{(2)} = {209 \, \kappa \over 2880 (16 \pi^2)^2} {1 \over
\epsilon}
\sqrt{-g} R^{\alpha \beta}_{~~\gamma \delta} R^{\gamma \delta}_{~~\rho
\sigma}
R^{\rho \sigma}_{~~
\alpha \beta}
\end{equation}

\noindent  The key feature is that higher powers of $R$ are involved at
higher
loops.  This is a
consequence of the structure of the energy expansion in a massless theory.
A
simple example can
illustrate the essentials of this
fact.  Consider a four graviton vertex, Fig. 1a.  Since each graviton brings a
factor of $\kappa$ (see Eq. 4 \, and recall that $\kappa \sim 1/M_{pl}$) the
Einstein
action gives
this a behavior

\begin{equation}
M_{Ein} \sim \kappa^2 p^2
\end{equation}

\noindent where $p$ is representative of the external momentum, whereas
the
Lagrangian at order $E^4$ have the behavior.

\begin{equation}
M_{H.O.} \sim c_1 \kappa^4 p^4
\end{equation}

\noindent If we use two of the Einstein vertices in a loop diagram, Fig. 1b,
the
momenta could be either external or internal, for example

\begin{equation}
M_{loop} \sim \int d^4l \, \kappa^2 l^2 {1 \over l^2} {1 \over (l-p)^2}
\kappa^2
l^2 \sim
\kappa^4 I(p) \sim \kappa^4 p^4
\end{equation}

\noindent If we imagine that the divergent integrals are regularized by
dimensional
regularization (which preserves the general covariance and which only
introduces
new scale dependent factors in logarithms not in powers), the Feynman
integral
must end up being expressed in terms of the external momenta.  Because no
masses
appear for gravitons, the momentum power of the final diagrams can be
obtained
easily by counting powers of $\kappa$.  The result is a divergence at order
$p^4$
and
can be absorbed into $c_1, c_2$.  Loop diagrams involve more gravitons
than
tree diagrams, hence more factors of $\kappa$, which by dimensional
analysis
means
that they are the same structure as higher order terms in the energy
expansion.

If we were to attempt a full phenomenological implementation of
gravitational
effective field theory
at one loop order, the procedure would be as follows:

\begin{enumerate}
\item Define the quantum degrees of freedom using the lowest order

$\left( {\cal O} (p^2) \right)$
effective Lagrangian, as done in Section 2.
\item Calculate the one loop corrections.
\item Combine the effects of the order $p^2$ and $p^4$ Lagrangians (given
earlier
in this section)
at tree level with the one loop corrections.  The divergences (and some
accompanying finite parts) of
the loop diagrams may be absorbed into renormalized coefficients of the
Lagrangian $(m, c_i, d_i)$, using some renormalization scheme which does
not
violate general
covariance.
\item Measure the unknown coefficients by comparison with some
experimental
measurement.
\item Having determined the parameters of the theory, one can make
predictions for
other
experimental observables, valid to ${\cal O} (p^4)$ in the energy
expansion.
\end{enumerate}

In practice the difficulty arises at step 4:  there is no observable that I am
aware of
which is
sensitive to reasonable values of any of the ${\cal O} (p^4)$
coefficients.  However the low
energy
content of the
gravitational effective field theory is not just contained in these
parameters. There is
a distinct class
of quantum corrections, uncovered in the above procedure, which are
independent
of the unknown
coefficients.  Moreover this class, the "leading quantum corrections", are
generally
dominant at
large distances over the other one loop gravitational corrections.  These are
discussed more fully in
the next section.

\section{Leading Quantum Corrections}

Although the ultraviolet behavior of quantum gravity has been heavily
studied to
learn about the behavior of general relativity as a fundamental theory, from
the
standpoint of effective field theory it is rather the infrared behavior which
is
more
interesting.  In the last section, the renormalization of the parameters in the
effective
action was described.  Although a technical necessity, this has no predictive
content.  However the low energy propagation of massless particles leads to
long
distance quantum corrections which are distinct from the effects of the local
effective Lagrangian.

A crucial distinction in this regard is whether the effective action may be
expanded
in a Taylor expansion in the momentum (or equivalently in powers of
derivatives).
If the result is analytic, it may be represented by a series of local
Lagrangians with
increasing powers of derivatives.  However nonanalytic effects cannot be
equivalent to local contributions, and hence are unmistakable signatures of
the low
energy particles.  Moreover, the nonanalytic effects can be dominant in
magnitude
over analytic corrections.  The expansion of the gravitational action is in
powers of
$q^2$ so that the first two terms of a matrix element will be

\begin{equation}
M_{local} = A \, q^2 \left( 1 + \alpha \kappa^2 q^2 + \ldots \right)
\end{equation}

\noindent As we will see in the next section, a graviton loop will have a
logarithmic
nonanalytic modification around flat space.

\begin{equation}
M_{full} = A \, q^2 \left( 1 + \alpha \kappa^2 q^2 + \beta \kappa^2 q^2 \,
ln \,
(-q^2) + \ldots \right)
\end{equation}

\noindent When massive matter fields are included in loops with gravitons
we may
also have nonanalytic terms of the form $m/ \sqrt{-q^2}$ instead of the
logarithm.[18]
Both of these effects have the property that they pick up imaginary
components for
timelike values of $q^2$ (i.e., for $q^2 > 0$ in this metric), as they are then
part of
the loop diagrams which are required for the unitarity of the $S$ matrix.
The
imaginary pieces arise from the rescattering of on-shell intermediate states,
and can
never contained in a local Lagrangian.  In addition, since $q^2$ can become
very
small, the nonanalytic pieces will satisfy $\mid ln \,(-q^2) \mid >> 1$ and
$\mid
m/\sqrt{-q^2} \mid << 1$ at low energy, thereby dominating over the
analytic
effect.  We can see that the nonanalytic terms have a distinct status as the
leading
quantum corrections due to long distance effects of massless particles.

The leading nonanalytic effects have the extra advantage that they involve
only the
massless degrees of freedom and the low energy couplings of the theory,
both of
which are known independent of the ultimate high energy theory.  The
massless
particles are the gravitons, photons and maybe neutrinos.  Only the lowest
energy
couplings are needed, since higher order effects at the vertices introduce
more
powers of $q^2$.  The low energy couplings are contained in the Einstein
action
and only depend on the gravitational  constant $G_N$.  So in distinction to
the
analytic contributions, which depend on the unknown parameters $c_1, c_2
\ldots$, the leading quantum corrections are parameter free.

Although our prime interest above has been the quantum corrections within
the
gravitational part of
the theory, we note that similar comments can be made if interactions other
than just
gravity are
present.  For example a theory with massless particles, such as photons in
QED,
can also generate
nonanalytic behavior in loop amplitudes when the photons are coupled to
gravity.
Let us call these
Class II nonanalytic corrections as compared to the Class I nonanalytic
effects found
due to the
quantum behavior of gravitons.  In addition there is a district type of
quantum
predictions (Class
III) which may also be predictions of the low energy theory once we allow
interactions in addition
to gravity.  This occurs for analytic terms in the energy expansion which are
accompanied by a
parameter with dimension $1/(mass)^n \; n \geq 1$.  The parameters $d_i$
in Eq. 21 \,
are examples.
The low energy theory can generate contributions to the parameter with
inverse
powers of a light
mass, while the Appelquist-Carazonne theorem[1] tells us that the effects of
a high
energy theory
would generally produce inverse powers of a heavy mass.  Therefore the
low
energy contribution
can be dominant, and the uncertainty caused by unknown high energy
theory is
minimal.  In the
case of the $d_i$ parameters, if the particles were strongly interacting QCD
would
generate a
gravitational charge radius corresponding to $d_i \approx 1/(1 GeV)^2$,
which
would be unlikely
to be changed by the underlying quantum theory of gravity.  Other examples
in the
case of QED
plus gravity have been worked out by Behrends and Gastmans.[18]
Classes II and
III corrections
(if
present) are most often larger than the gravitational leading quantum
corrections
(Class I) because
their form need not be expansions in the small quantity $G_N$.  [An
exception is
the gravitational
potential at large distance, where analytic corrections have no effect on the
$1/r^3$
term.]

\section{Example:  The Gravitational Potential}

In this section I describe in detail an example which demonstrates the
extraction of
the leading
quantum corrections.  The gravitational interaction of two heavy objects
close to
rest described in
lowest order by the Newtonian potential energy

\begin{equation}
V(r) = -{G m_1 m_2 \over r}
\end{equation}

\noindent This is modified in general relativity by higher order effects in
$v^2/c^2$
and by
nonlinear terms in the field equations of order ${Gm \over rc^2}$ (which
are of the
same order as
$v^2/c^2$).  While a simple potential is not an ideal relativistic concept, the
general
corrections
would be of the form[7]

\begin{equation}
V(r) =- {G m_1 m_2 \over r} \left[ 1 + a {G (m_1 + m_2 ) \over r c^2} +
\dots
\right]
\end{equation}

\noindent The number "a" would depend on the precise definition of the
potential
and would be
calculable in the Post-Newtonian expansion.  At some level there will be
quantum
corrections also.
By dimensional analysis, we can figure out the form that these should take.
Since
they arise from
loop diagrams, they will involve an extra power of $\kappa^2 \sim G$, and
if they
are quantum
corrections they will be at least linear in $\hbar$.  If the effects are due to
long range
propagation
of massless particles, the other dimensionful parameter is the distance $r$.
The
combination

\begin{equation}
{G \hbar \over r^2 c^3}
\end{equation}

\noindent is dimensionless and provides an expansion parameter for the
long
distance quantum
effects.  We then expect a modification to the potential of the form

\begin{equation}
V(r) = -{G m_1 m_2 \over r} \left[ 1 + a {G (m_1 + m_2) \over r c^2} + b
{G
\hbar \over r^2
c^3}
+ \ldots \right]
\end{equation}

\noindent and our goal is to calculate the number b for an appropriate
definition of
the potential.[5]

The Newtonian potential can be found as the nonrelativistic limit of graviton
exchange, Fig. 2.  In
the harmonic gauge, the graviton propagator is

\begin{equation}
i D_{\mu \nu \alpha \beta} (q) = {i \over q^2 - i \epsilon} P_{\mu \nu,
\alpha \beta}
\end{equation}

\noindent with

\begin{equation}
P_{\mu \nu, \alpha \beta} = {1 \over 2} \left[ \eta_{\mu \alpha} \eta_{\nu
\beta} +
\eta_{\nu
\alpha} \eta_{\mu \beta} - \eta_{\mu \nu} \eta_{\alpha \beta} \right]
\end{equation}

\noindent with $\eta_{\mu \nu}$ being the flat space matrix $\eta_{\mu \nu}
= diag
(1, -1, -1,
-1)$.  The matter stress tensor has the on-shell matrix element

\begin{equation}
V_{0 \mu \nu} (q) \equiv < p' \mid T_{\mu \nu} \mid p > = p_{\mu}
p'_{\nu} +
p'_{\mu}
p_{\nu} +
{1 \over 2} q^2 \eta_{\mu \nu}
\end{equation}

\noindent in our normalization convention

\begin{equation}
<p' \mid p > = 2E (2 \pi)^3 \delta^3 (p - p')
\end{equation}

\noindent [Here the subscript 0 indicates that this form holds before the
inclusion of
radiative
corrections.]  Graviton exchange then yields

\begin{equation}
M_{12} = {\kappa^2 \over 4} V^{(1)}_{0 \mu \nu} (q) D^{\mu \nu,
\alpha \beta}
(q)
V^{(2)}_{0
\alpha \beta} (-q).
\end{equation}

The static limit corresponds to $q_{\mu} = (0, {\bf q})_{\mu}$ and

\begin{equation}
{1 \over 2m_1} V^{(1)}_{\mu \nu} (q) = m_1 \delta_{\mu 0} \delta_{\nu
0}
\end{equation}

\noindent where the ${1 \over 2m_1}$ accounts for the covariant
normalization
factor.  The
Newtonian potential is then found from the Fourier transform of

\begin{equation}
{1 \over 2m_1 2m_2} M_{12} \sim -{\kappa^2 \over 8} {m_1 m_2 \over {\bf
q}^2}
\end{equation}

\noindent which in coordinate space yields

\begin{equation}
V(r) = -\int {d^3 q \over (2 \pi)^3} e^{i{\bf q} \cdot {\bf r}} {\kappa^2
\over
8}{m_1 m_2 \over
{\bf q}^2} = - G {m_1 m_2 \over r}.
\end{equation}

\noindent Of course, despite the description of quantizing, gauge fixing etc.,
this is
purely a
classical result.

In order to define a quantum potential one can consider the set of one
particle
reducible graphs of
Fig. 3, where the heavy dots signify the full set of radiative corrections to
the vertex
function and
the graviton propagator.  These corrections are given explicitly in Fig. 4, 5.
It is
this set which we
will examine.  Fortunately we will be able to extract the information that we
need
for the vacuum
polarization from the work of others.  This leaves the vertex correction to be
worked out here.

The vertices required for the calculation follow from the Lagrangians given
previously.  For the
vertices pictured in Fig. 6, we find

\begin{equation}
Fig. \, 6a \, \qquad \tau_{\mu \nu} (p, p') = {-i \kappa \over 2}
\left( p_{\mu}
p'_{\nu} + p'_{\mu}
p_{\mu} -
g_{\mu
\nu} \left[ p \cdot p' - m^2 \right] \right)
\end{equation}

\noindent and

\begin{eqnarray}
Fig. \, 6b \, \qquad V_{\eta \lambda, \rho \sigma} & = & {i \kappa^2 \over
2}
\left\{ I_{\eta
\lambda, \alpha
\delta}
I_{~\beta, \rho \sigma}^{\delta} \left( p^{\alpha} p'^{\beta} + p^{\alpha}
p'^{\beta} \right)
\right. \nonumber \\
& & - {1 \over 2} \left( \eta_{\eta \lambda} I_{\rho \sigma, \alpha \beta} +
\eta_{\rho
\sigma} I_{\eta  \lambda, \alpha \beta} \right) p'^{\alpha} p^{\beta}
\nonumber \\
& & \left. - {1 \over 2} \left( I_{\eta \lambda, \rho \sigma} - {1 \over 2}
\eta_{\eta
\lambda}
\eta_{\rho \sigma} \right) \left[ p \cdot p' - m^2 \right] \right\}
\end{eqnarray}

\noindent where

\begin{equation}
I_{\alpha \beta, \gamma \delta} \equiv {1 \over 2} \left( \eta_{\alpha
\gamma}
\eta_{\beta \delta} +
\eta_{\alpha \delta} \eta_{\beta \gamma} \right)
\end{equation}

\noindent The graviton vertex is found most easily by using Eq. 8 plus
Eq. 5 with the
background metric being
expanded as $\bar{g} (x) = \eta_{\mu \nu} + \kappa H^{ext}_{\mu \nu}
(x)$,
where we pick out
the
vertex with one external field and two quantum fields.  After some work,
this can
be put into the
form
\begin{eqnarray}
\tau^{\mu \nu}_{\alpha \beta, \gamma \delta} &=& {i \kappa \over 2} \left(
P_{\alpha \beta,
\gamma
\delta} \left[ k^{\mu} k^{\nu} + (k - q)^{\mu} (k - q)^{\nu} + q^{\mu}
q^{\nu} -
{3 \over 2}
\eta^{\mu \nu} q^2 \right] \right. \nonumber \\
& & + 2q_{\lambda} q_{\sigma} \left[ I_{~~~\alpha \beta}^{\lambda \sigma,}
I_{~~~\gamma
\delta}^{\mu \nu,} + I_{~~~\gamma \delta}^{\lambda \sigma,} I_{~~~\alpha
\beta}^{\mu \nu,} -
I_{~~~\alpha \beta}^{\lambda \mu,} I_{~~~\gamma \delta}^{\sigma \nu,} -
I_{~~~\alpha
\beta}^{\sigma \nu,}
I_{~~~\gamma \delta}^{\lambda \mu,} \right] \nonumber \\
& &+ \left[ q_{\lambda} q^{\mu} \left( \eta_{\alpha \beta} I_{~~~\gamma
\delta}^{\lambda \nu,} +
\eta_{\gamma \delta} I_{~~~\alpha \beta}^{\lambda \nu,} \right) + q_{\lambda}
q^{\nu} \left(
\eta_{\alpha \beta} I_{~~~\gamma \delta}^{\lambda \mu,} + \eta_{\gamma
\delta}
I_{~~~\alpha
\beta}^{\lambda \mu,} \right) \right. \nonumber \\
& &\left. - q^2 \left( \eta_{\alpha \beta} I_{~~~\gamma \delta}^{\mu \nu,} +
\eta_{\gamma \delta}
I_{~~~\alpha \beta}^{\mu \nu,} \right) - \eta^{\mu \nu} q^{\lambda}
q^{\sigma} \left(
\eta_{\alpha
\beta} I_{\gamma \delta, \lambda \sigma} + \eta_{\gamma \delta} I_{\alpha
\beta,
\lambda \sigma}
\right) \right] \nonumber \\
& & + \left[ 2 q^{\lambda} \left( I_{~~~\alpha \beta}^{\sigma \nu,} I_{\gamma
\delta,
\lambda
\sigma} (k - q)^{\mu} + I_{~~~\alpha \beta}^{\sigma \mu,} I_{\gamma \delta,
\lambda
\sigma} (k -
q)^{\nu} \right. \right. \nonumber \\
& & \left. - I_{~~~\gamma \delta}^{\sigma \nu,} I_{\alpha \beta, \lambda
\sigma}
k^{\mu} -
I_{~~~\gamma \delta}^{\sigma \mu,} I_{\alpha \beta, \lambda \sigma} k^{\nu}
\right)
\nonumber \\
& & \left. + q^2 \left( I_{~~~\alpha \beta}^{\sigma \mu,} I_{\gamma \delta,
\sigma}^{~~~~\nu} +
I_{\alpha \beta, \sigma}^{~~~~\nu} I_{~~\alpha \delta}^{\sigma \mu,} \right) +
\eta^{\mu
\nu}
q^{\lambda} q_{\sigma} \left( I_{\alpha \beta, \lambda \rho} I_{~~\gamma
\delta}^{\rho \sigma,} +
I_{\gamma \delta, \lambda \rho} I_{~~~\alpha \beta}^{\rho \sigma,} \right)
\right]
\nonumber \\
& & + \left\{ \left( k^2 + (k-q)^2 \right) \left( I_{~~~\alpha \beta}^{\sigma
\mu,}
I_{\gamma \delta,
\sigma}^{~~~~\nu} + I_{~~~\alpha \beta}^{\sigma \nu,} I_{\gamma \delta,
\sigma}^{~~~~\mu}
- {1 \over 2}
\eta^{\mu \nu} P_{\alpha \beta, \gamma \delta} \right) \right. \nonumber \\
& & \left. \left. - \left( k^2 \eta_{\gamma \delta} I_{~~~\alpha \beta}^{\mu
\nu,} + (k -
q)^2
\eta_{\alpha \beta} I_{~~\gamma \delta}^{ \mu \nu,} \right) \right\} \right)
\end{eqnarray}

\noindent with $I_{\alpha \beta, \gamma \delta}$ defined in Eq. 52 and
$P_{\alpha
\beta, \gamma
\delta}$ defined in Eq. 42.

The diagrams involved for the vertex correction are given in Fig. 4.  We
will argue
below that Fig.
4d,e,f will not contribute to the
leading quantum corrections, so that we need to
calculate only
diagrams 4b and 4c.  These have the form

\begin{eqnarray}
M^{\mu \nu}(3b) & = & \int {d^4 k \over (2 \pi)^4} i \tau_{\eta \lambda}
(p, p' -
k) {i \over (k -
p')^2
-  m^2 + i \epsilon} i \tau_{\rho \sigma} (p' - k,p') \nonumber \\
& & i D^{\eta \lambda, \alpha \beta} (k - q) i \tau^{\mu \nu}_{\alpha \beta,
\gamma
\delta}
i D^{\gamma
\delta, \rho \sigma} (k) \nonumber \\
M^{\mu \nu}(3c) & = & \int {d^4 k \over (2 \pi)^4} i V_{\eta \lambda, \rho
\sigma} i D^{\eta
\lambda,
\alpha \beta} (k - q) i \tau^{\mu \nu}_{\alpha \beta, \gamma \delta} i
D^{\gamma
\delta, \rho
\sigma} (k)
\end{eqnarray}

Before proceeding, it is worth examining the structure of the answer.  The
general
vertex may be
described by two form factors

\begin{eqnarray}
V_{\mu \nu} \equiv <p' \mid T_{\mu \nu} \mid p > & = & F_{1} (q^2)
\left[
p_{\mu} p'_{\nu} +
p'_{\mu} p_{\nu} + q^2 {\eta_{\mu \nu} \over 2} \right] \nonumber \\
& & + F_2 (q^2) \left[ q_{\mu} q_{\nu} - g_{\mu \nu} q^2 \right]
\end{eqnarray}

\noindent with normalization condition $F_1 (0) = 1$.  The expansion in
energy
corresponds to an
expansion of the form factors in powers of $q^2$.  The one loop diagrams
of Fig.
6 have an extra
power of $\kappa^2$ compared to the tree level vertex, and $\kappa^2$ has
dimensions
$(mass)^{-2}$, so
that $\kappa^2 m^2$ and $\kappa^2 q^2$ form dimensionless
combinations.
However loop
diagrams will
also produce non analytic terms with the form $ln (-q^2)$ and $\sqrt{{m^2
\over
-q^2}}$, which
also are dimensionless.  Also contributing to the form factors are the terms
in the
higher order
Lagrangian as these give extra factors of $q^2$.  By working out these
contributions and taking
the general form of the loop diagrams from dimensional considerations we
obtain
form factors

\begin{eqnarray}
F_1 (q^2) & = & 1 + d_1 q^2 + \kappa^2 q^2 \left( \ell_1 + \ell_2 ln {(-
q^2) \over
\mu^2} +
\ell_3 \sqrt{{m^2 \over -q^2}} \right) + \ldots \nonumber \\
F_2 (q^2) & = & -4 (d_2 + d_3) m^2 +  \kappa^2 m^2 \left( \ell_4 + \ell_5
ln {(-
q^2) \over
\mu^2} +
\ell_6 \sqrt{{m^2 \over -q^2}} \right) + \ldots
\end{eqnarray}

\noindent where $\ell_i (i = 1, 2 \ldots 6)$ are numbers which come from
the
computation of the
loop diagrams.  There can be no corrections in $F_1 (q^2)$ of the form
$\kappa^2
m^2$ because of the
normalization condition $F_1 (0) = 1$.  The constant $\mu^2$ can be
chosen
arbitrarily, with a
corresponding shift in the constants $\ell_1$ and $\ell_4$.  The ellipses
denote
higher powers of
$q^2$.  The constants $\ell_1$ and $\ell_4$ will in general be divergent,
while
$\ell_2, \ell_3,
\ell_5$ and $\ell_6$ must be finite.  For $q^2$ time-like, $ln (-q^2)$ and
$\sqrt{{m^2 \over
-q^2}}$ pick up imaginary parts which correspond to the physical (on
shell)
intermediate states as
described by unitarity.  Recall that $d_i$ represents the unknown effects of
the true
high energy
theory, while $\ell_1$ and $\ell_4$ come largely from the high energy end
of the
loop integrals.
For these high energies we have no way of knowing if the loop integrals are
well
represented by
the
low energy vertices and low energy degrees of freedom--almost certainly
they are
not.  Therefore it
is logical, as well as technically feasible, to combine $\ell_1$ and $\ell_4$
with the
constants
$d_i$, producing renormalized values

\begin{eqnarray}
d^{(r)}_1 (\mu^2) & = & d_1 + \kappa^2 \ell_1 \nonumber \\
d^{(r)}_2 (\mu^2) + d^{(r)}_3 (\mu^2) & = & d_1 + d_3 - \kappa^2
{\ell_4 \over
4}
\end{eqnarray}

\noindent It is these renormalized values which would be (in principle)
measured by
experiment,
and the $\mu^2$ labeling $d_i^{(r)} (\mu^2)$ indicates that the measured
value
would depend on
the choice of $\mu^2$ in the logarithms, although all physics would be
independent
of $\mu^2$.

In forming a gravitational interaction of two particles, one combines the
vertices
with the
propagator.  Temporarily, leaving aside the vacuum polarization, one has

\begin{eqnarray}
{\kappa^2 \over 4} && V^{(1)}_{\mu \nu} (q) D^{\mu \nu, \alpha \beta}
(q)
V^{(2)}_{\alpha
\beta} (-q)
\nonumber \\
&&= {\kappa^2 \over 2q^2} \left[ F^{(1)}_1 (q^2) F^{(2)}_1 (q^2) \left\{
p_1
\cdot p_2 p_1'
\cdot
p_2' + p_1 \cdot p_2' p_2 \cdot p_1' - m^2_1 m^2_2 \right\} \right.
\nonumber \\
&& \left. + {q^2 \over 2} \left\{ F^{(1)}_1 (q^2) F^{(2)}_2 (q^2) m^2_1
+
F^{(2)}_1 (q^2)
F^{(1)}_2 (q^2) m^2_2 \right\} \right] \nonumber \\
&& \approx {\kappa^2 m^2_1 m^2_2 \over 2} \left\{ {1 \over q^2} + 2\left(
d_1 - 2 d_2 - 2 d_3 \right)  \right. \nonumber\\
&&\left. + \kappa^2 \left( (2\ell_1 - \ell_4) +
(2 \ell_2 - \ell_5) ln
\left( {-q^2 \over \mu^2} \right) + (2\ell_3 - \ell_6)\sqrt{{m^2
\over -q^2}} \right)
\right\}
\end{eqnarray}

\noindent where the second line is the approximate result in the static limit.
Linear
analytic terms in
$q^2$ in the form factors yield {\em constants} in the interactions, which in
turn
correspond to a
point (delta function) interaction

\begin{equation}
\int {d^3 q \over (2 \pi)^3} e^{-i \vec{q} \cdot \vec{r}} = \delta^3 (x)
\end{equation}

\noindent The non analytic terms however lead to {\em power law}
behavior since

\begin{eqnarray}
\int {d^3q \over (2 \pi)^3} e^{-i \vec{q} \cdot \vec{r}} {1 \over q} = {1
\over 2
\pi^2 r^2}
\nonumber \\
\int {d^3q \over (2 \pi)^3} e^{-i \vec{q} \cdot \vec{r}} ln \vec{q}^2 = {-1
\over 2
\pi^2 r^3}
\end{eqnarray}

\noindent Therefore, the long distance corrections to the gravitational
interactions
come uniquely
from the non analytic terms in the loop diagrams.

A similar result holds for the vacuum polarization diagram.  If we
temporarily
suppress the Lorentz
indices and relative constants of order unity, the generic form of the vacuum
polarization follows
from dimensional counting

\begin{equation}
\pi (q^2) = \kappa^2 q^4 \left[ c_1 + c_2 + \ell_7 + \ell_8 ln (-q^2) \right]
\end{equation}

\noindent such that the propagator is modified as

\begin{equation}
{1 \over q^2} + {1 \over q^2} \pi (q^2) {1 \over q^2} + \ldots = \left\{ {1
\over
q^2} + \kappa^2
\left[
c_1 + c_2 + \ell_7 + \ell_8 ln (-q^2) \right] \right\}
\end{equation}

\noindent In these formulas $c_1$ and $c_2$ are the unknown parameters
from the
higher order
Lagrangian of Eq. 20, and $\ell_7, \ell_8$ are constants calculable in the
vacuum
polarization
diagram.  Again $\ell_7$ is divergent, but the combination $(c_1 + c_2 +
\ell_7)$
forms a
renormalized parameter which could in principle be measured.  As above,
constants
in this
propagator lead to a $\delta^3(x)$ interaction, while the logarithm
corresponds to a
long range
effect.

By focusing only on the non-analytic terms, we simplify the calculation
somewhat.
These are
independent of the regularization scheme.  The non-analytic pieces of the
relevant
Feynman integrals are given in the Appendix.  We note that there are several
useful
properties that
we can exploit.  For example, any factor of $k^2$ or $(k-q)^2$ in the
numerator
automatically
removes any non-analytic behavior.  For example

\begin{eqnarray}
\int {d^4 k \over (2 \pi)^4} {1 \over k^2 (k-q^2)} {k^2 \over (k-p')^2 -
m^2} =
\int {d^4 k
\over (2 \pi)^4} {1 \over (k-q)^2} {1 \over (k-p')^2 - m^2} \nonumber \\
= \int {d^4 k' \over (2 \pi)^4} {1 \over k'^2} {1 \over (k' -p)^2 - m^2} =
I(p^2)
\end{eqnarray}

\noindent Where in the second line we have shifted to $k' = k - q$.  The
result is a
function of
$m^2$ only and has no $q^2$ dependence.  Thus the non-analytic terms
vanish for
any integrands
which would vanish on shell gravitons.  This is a reflection of the Cutkosky
rules
and is not
surprising as the non analytic terms accompany imaginary parts in the
amplitudes,
and these latter
could be seen using on shell states and unitarity.  As a result, the vertex
function
simplifies slightly as
all the components in the curly brackets \{\} of Eq. 53 do not contribute.
Also,
factors
of $k \cdot q$
may be written as $2k \cdot q = k^2 - (k - q)^2 + q^2 \rightarrow q^2$ in
the loop
integrals.  With
these results, it is not hard to show that

\begin{equation}
q_{\mu} \tau^{\mu \nu}_{\alpha \beta, \gamma \delta} = 0
\end{equation}

\noindent in any diagram, where in fact the vanishing of the results occurs
individually for each of
the terms in the square brackets of Eq. 53. These conditions provide a set of
checks on
the
calculation that were found to be useful.

The calculation of the nonanalytic terms is straightforward although a bit
tedious
due to the lengthy
form of the triple graviton coupling.  For the diagram of Fig. 4b, I find

\begin{eqnarray}
\Delta F_1 (q^2) & = & {\kappa^2 q^2 \over 32 \pi^2} \left\{ \left[ {1 \over
4} - 2
+ 1 + 0 \right]
ln (-
q^2) + \left[ {1 \over 16} - 1 + 1 + 0 \right] {\pi^2 m \over \sqrt{-q^2}}
\right\}
\nonumber \\
& = & {\kappa^2 q^2 \over 32 \pi^2} \left\{ -{3 \over 4} ln (-q^2) + {1
\over 16}
{\pi^2 m \over
\sqrt{-
q^2}} \right\} \nonumber \\
\Delta F_2 & = & {\kappa^2 m^2 \over 32 \pi^2} \left\{ \left[ 1 - 3 + 8 - 3
\right] ln
(-q^2) + \left[
{7  \over 8} - 1 + 2 - 1 \right] {\pi^2 m \over \sqrt{-q^2}} \right\}
\nonumber \\
& = & {\kappa^2 m^2 \over 32 \pi^2} \left\{ 3 ln (-q^2) + {7 \over 8}
{\pi^2 m^2
\over \sqrt{-
q^2}}
\right\}
\end{eqnarray}

\noindent where the sequence of numbers in the first version of $F_i$ refers
to the
four sets of
terms in square brackets in Eq. 53 respectively.  For Fig. 4c, I obtain

\begin{eqnarray}
\Delta F_1 &=&{\kappa^2 q^2 \over 32 \pi^2} \left[ 0 + 2 + 0 - 2 \right]
ln(-q^2)
\nonumber \\
&=& 0 \\ \nonumber
\Delta F_2 &=& {\kappa^2 m^2 \over 32 \pi^2} \left[ -{25 \over 3} + 0 + 2
+ 2
\right] ln (-q^2)
\nonumber \\
&=& {\kappa^2 m^2 \over 32 \pi^2} \left[ - {13 \over 3} ln (-q^2) \right] .
\end{eqnarray}

\noindent The diagrams of Fig. 4d,e,f do not have any non-analytic terms of
the form
considered
here because the matter fields are massive or the loop integrals are
independent of $q^2$.  Fig. 4d does have an infrared
divergence similar to that
of the vertex correction in QED.  This can be handled in the same fashion as
the
QED case,[20]
i.e., by including soft radiative effects.  If we were to actually attempt to
apply the
quantum
potential phenomenologically, it would be important to include such effects.
However for our
purposes we do not need to consider them further.  The resulting non-
analytic
contributions to
$F_1, F_2$ are then

\begin{eqnarray}
F_1(q^2) &=& 1 + {\kappa^2 \over 32 \pi^2} q^2 \left[ - {3 \over 4} ln (-
q^2) +
{1 \over 16}
{\pi^2
m \over \sqrt{-q^2}} \right] \nonumber \\
F_2(q^2) &=& {\kappa^2 m^2 \over 32 \pi^2} \left[ - {4 \over 3} ln(-q^2)
+ {7
\over 8} {\pi^2 m
\over \sqrt{-q^2}} \right]
\end{eqnarray}

The non analytic terms in the vacuum polarization can be obtained by the
following
procedure.
The divergent parts of the vacuum polarization have been calculated, by
't~Hooft
and Veltman[3]
employing the same gauge fixing scheme as used in the present paper, using
dimensional
regularization.  When only massless particles appear in the diagram, the $ln
(-q^2)$
terms can be
read off of the coefficients of $1/(d-4)$ using a relatively well known trick.
The
vacuum
polarization graph has dimension of $(mass)^2$ and will be calculated from
Feynman integrals of
the following form

\begin{eqnarray}
I(q^2) &=& \kappa^2 \mu^{4-d} \int {d^d k \over (2 \pi)^d} f(k, q)
\nonumber \\
&=& \kappa^2 q^4 \left( {\mu \over q} \right)^{4-d} \left[ {a \over d-4} +
b \right]
\end{eqnarray}

\noindent for some integrand $f(k, q)$.  The arbitrary scale factor $\mu$,
with
dimension of
$(mass)^1$, has been inserted in order to maintain the proper dimension for
the
overall integral.
The second form then follows uniquely from dimensional analysis (since
$q$ is the
only other
dimensionful quantity), where $a$ and $b$ are constants which may depend
on $d$
but do not
contain further poles as $d \rightarrow 4$.  Since

\begin{eqnarray}
{1 \over d-4} \left( {\mu \over q} \right)^{4-d} & = & {1 \over d-4}
e^{{d-4 \over
2}ln \left(
{q^2  \over \mu^2} \right)} \, \qquad (a) \nonumber \\
& = & {1 \over d-4} + {1 \over 2} ln \left( {q^2 \over \mu^2} \right) +
O(d-4)
{}~~~(b)
\end{eqnarray}

\noindent the logarithm will always share the coefficient of ${1 \over d-4}$
in the
combination
given in Eq. 70(b).  't~Hooft and Veltman find that the divergent part of the
graviton
plus ghost
vacuum polarization diagrams is equivalent to the Lagrangian in Eq. 30.  This
is therefore
equivalent to the
gravitational logs in the diagram being given by

\begin{eqnarray}
\Pi_{\alpha \beta, \gamma \delta} & = & - {\kappa^2 \over 32 \pi^2} ln (-
q^2)
\left\{ {21 \over
120}
q^4  I_{\alpha \beta \gamma \delta} + {23 \over 120} q^4 \eta_{\alpha
\beta}
\eta_{\gamma \delta}
\right.  \nonumber \\
& & - {23 \over 120} \left( \eta_{\alpha \beta} q_{\gamma} q_{\delta} +
\eta_{\gamma \delta}
q_{\alpha} q_{\beta} \right) \nonumber \\
& & + {21 \over 240} \left( q_{\alpha} q_{\delta} \eta_{\beta \gamma} +
q_{\alpha} q_{\gamma}
\eta_{\beta \delta} + q_{\beta} q_{\gamma} \eta_{\alpha \delta} + q_{\beta}
q_{\delta}
\eta_{\alpha \gamma} \right) \nonumber \\
& & \left. + {11 \over 30} q^{\alpha} q^{\beta} q^{\gamma} q^{\delta}
\right\} +
(non logs)
\end{eqnarray}

\noindent When we construct the potential, a related form will be needed

\begin{eqnarray}
P_{\mu \nu, \alpha \beta} \Pi^{\alpha \beta, \gamma \delta} P_{\gamma
\delta, \rho
\sigma} =
{\kappa^2 q^4 \over 32 \pi^2} \left[ {21 \over 120} \left( \eta_{\mu \rho}
\eta_{\nu
\sigma} +
\eta_{\nu
\rho} \eta_{\mu \sigma} \right) + {1 \over 120} \eta_{\mu \nu}
\eta_{\rho \sigma} \right] \nonumber \\
\left[ -ln  (-q^2) \right] + \ldots
\end{eqnarray}

\noindent where all terms involving $q_{\mu}, q_{\rho}$ etc. can be
dropped since
$q_{\mu}$
contracted with the vertex function gives a vanishing result.

If we combine these diagrams as shown in Fig. 3 in order to form a one-
particle-
reducible
gravitational interaction, we find

\begin{eqnarray}
& - & {\kappa^2 \over 4} {1 \over 2m_1} V^{(1)}_{\mu \nu} (q) \left[ i
D^{\mu
\nu, \alpha
\beta} (q) + i
D^{\mu \nu, \rho \sigma} i \Pi_{\rho \sigma, \eta \lambda} i D^{\eta
\lambda,
\alpha \beta} \right]
V_{\alpha \beta} (q) {1 \over 2m_2} \nonumber \\
& \approx & 4 \pi G m_1 m_2 \left[ {i \over {\bf q}^2} - {i \kappa^2 \over
32
\pi^2} \left[ - {127
\over
60}  ln ({\bf q}^2) + {\pi^2 (m_1 + m_2) \over 2 \sqrt{{\bf q}^2}} \right]
+ const
\right]
\end{eqnarray}

\noindent where in the second line we have taken the nonrelativistic limit.
This can
be converted
into the coordinate space form by Fourier transforming, with the result

\begin{equation}
V(r) = - {Gm_1 m_2 \over r} \left[ 1 - {G(m_1 + m_2) \over r c^2} - {127
\over
30 \pi^2} {G
\hbar \over r^2 c^3} \right]
\end{equation}

\noindent This conforms to the general structure of the gravitational
interactions
given in Eq. 41.

One interesting consequence of the quantum correction is that it appears that
there is
no such thing
as a purely classical source for gravity.  In electrodynamics, the corrections
to the
vertex function
are such that as one takes the particles' mass to infinity all quantum
corrections will
vanish.
Therefore by taking the $m \rightarrow \infty$ limit to the full theory, one
obtains a
purely classical
source for electromagnetism.  In the case of gravity this does not work, as
the
quantum corrections
remain even as $m \rightarrow \infty$ and in fact share the same dependence
on
$m$ as does the
classical theory.  This is because the gravitational coupling itself grows with
$m$
so that the
expected decrease of quantum effects as $m \rightarrow \infty$ is
compensated for
by the increased
coupling constant.

\section{The Extreme Low Energy Limit}

The gravitational effective field theory appears to behave normally over
distance/energy scales where we have experience with gravity and quantum
theory.
We have imagined quantizing in a very large box in order to exclude
wavelengths of
the scale
of the Universe, and the effective field theory methodology separates out the
high
energy effects.  We expect modifications to the theory on the high energy
side.  Of
significant interest is whether there is a fundamental incompatibility between
gravity
and quantum mechanics at the extreme infrared side if we remove the low
energy
cut off.  In this section, I briefly discuss the reasons for suspecting that
there might
be a
problem, although I do not resolve the issues.

Gravitational effects can build up over long distances/times.  Most
disturbing in this
regard are the singularity theorems of Hawking and Penrose[21] which state
loosely
that a matter distribution evolved in space time by the Einstein action almost
always
has a true singularity in either the past or future.  [The exact assumptions
are
more
precisely stated in the original works, but apply to essentially all situations
that we
care about although simple cases such as Minkowski space or a single stable
star
are exceptions.]  Thus while it may be possible to locally specify a smooth
set of
coordinates with a small curvature, if we try to extend this condition to the
whole
space-time manifold using the order $E^2$ Einstein equations, there will be
at least
one location in the distant past or future where the curvature becomes
singular.  The
Big Bang in our standard cosmology is an example.  From the standpoint of
effective field theory, it is not the singularity itself which is the concern.
Once the
curvature becomes large, the $R^2, R^3$ etc. terms in the action becomes
important and the evolution is different than is assumed in the derivation of
the
theorem.  There is no longer any indication that a true singularity must
develop
beyond this scale.  However it {\em is} bothersome that the curvature must
get
large.  For gravitational effective field theory, the content of the
singularity
theorems is that it is difficult to specify a space-time where the curvature is
everywhere small.  The gravitational effective field theory would work over
scales
which have
small curvatures, but would not be able to work on scales which encompass
the
putative
singularity.  However, it may be possible to modify the effective field
theory
treatment to include
the singular region as an extended gravitational source, much as solitonic
Skyrmions can be treated
as a heavy chiral source in chiral perturbation theory.

Another global potential problem in the extreme infrared is the existence of
horizons.  Black
holes can exist which have their horizon in low curvature regions.  This is
not a
problem in a local patch near the horizon, but may lead to difficulties if we
extend
the region to all of space-time.  Quantum information that would have
otherwise
flowed to infinity disappears into the black hole.  Quantum coherence is lost
on the
largest time scales.  It is not clear that this is a contradiction with the
ultimate
quantum theory, but
at the least we end up with a situation which is beyond our experience in
physics and for which we are unsure as to how to apply quantum ideas.

\section{Summary}

Besides being a useful calculational tool, effective field theory provides a
good way
to think about the different energy scales of a theory.  In the case of
quantum
gravity it allows one to separate the effects of the unknown high energy
theory from
the known degrees of freedom at low energy.  Gravity and quantum
mechanics
seem to be compatible over a large range of energies corresponding
to our range of experience in physics.  In this range quantum predictions
can be
extracted from
generally covariant theories in the same way as done in other effective field
theories.

A particular class of one loop effects have been isolated and shown to give
the
leading quantum correction in an expansion in the energy or inverse
distance.
These nonanalytic terms come uniquely from the long distance propagation
of the
massless particles.  The example of the leading corrections to the vertex and
vacuum polarization diagrams in flat space has been discussed in detail, and
these
have been combined to form an effective potential.

The class of leading quantum corrections is not optional.  The set of
intermediate
states is a known consequence of the low energy theory.  It will also be
found in a
full quantum gravity theory as long as our universe is reproduced in the low
energy
limit.

\begin{center}
{\bf Acknowledgement}
\end{center}

I would like to thank Jennie Traschen and David Kastor for numerous
discussions
on this topic
and G. Esposito-Farese, S. Deser, H. Dykstra, E. Golowich, B. Holstein,
G.
Leibbrandt, and J.
Simon for useful comments.

\section{Appendix}

Below are the non-analytic terms which arise in many Feynman integrals
needed in
Section 5.  These are given the "mostly negative" metric $q^2 = q^2_0 -
{\bf
q}^2$.  I use the
abbreviations $L = ln \, (-q^2), S = \pi^2 m / \sqrt{-q^2}$.

\begin{eqnarray}
J & = & \int {d^4 k \over (2 \pi)^4 } \: {1 \over k^2 (k-q)^2} = {i \over 32
\pi^2} \left[ -2L \right] + \ldots \nonumber \\
J_{\mu} & = & \int {d^4k \over (2 \pi )^4} \: {k_{\mu} \over k^2 (k -
q^2)} = {i
\over
32 \pi^2} q_{\mu} \left[ -L \right] + \ldots \nonumber \\
J_{\mu \nu} & = & \int {d^4k \over (2 \pi )^4} {k_{\mu} k_{\nu} \over
k^2 (k-
q^2)} =
{i \over 32 \pi^2} \: \left[ q_{\mu} q_{\nu} \left\{ - {2 \over 3} L \right\} -
g_{\mu
\nu} \, q^2 \left\{ - {1 \over 6} L \right\} \right] \nonumber \\
& &+ \ldots
\end{eqnarray}

\noindent In the following, the external momentum $p'$ is on shell as is $p
= p' -
q$.

\begin{eqnarray}
I & = & \int {d^4 k \over (2 \pi)^4} {1 \over k^2} {1 \over (k-q)^2} {1
\over
[(k-p')^2  - m^2]} = {i \over 32 \pi^2 m^2} [ - L - S ] \nonumber \\
I_{\mu} & = & \int {d^4 k \over (2 \pi)^4} {k_{\mu} \over k^2 (k - q)^2
[(k -
p')^2 -
m^2]} = {i \over 32 \pi^2 m^2} \left[ p'_{\mu} \left\{ \left( 1 + {q^2 \over
2m^2}
\right) L \right. \right. \nonumber \\
& & \left. \left.+ {1 \over 4} {q^2 \over m^2} S \right\}+ q_{\mu} \left\{ -
L - {1
\over
2}S \right\} \right] \nonumber \\
I_{\mu \nu} & = & \int {d^4 k \over (2 \pi)^4} {k_{\mu}
k_{\nu} \over k^2 (k - q)^2  \left[(k - p')^2 + m^2 \right]} = {i \over 32
\pi^2
m^2}
\left[ p'_{\mu} p'_{\nu} \left\{ -{q^2 \over 2m^2} L \right. \right.
\nonumber \\
& &\left. - {q^2 \over 8m^2} S \right\} + (p'_{\mu} q_{\nu} + p'_{\nu}
q_{\mu}) \left\{
{1 \over 2} \left( 1 + {q^2 \over  m^2}\right) L + {3 \over 16} {q^2 \over
m^2}
S \right\}\nonumber \\
& &\left. + q_{\mu} q_{\nu} \left\{ - L - {3 \over 8} S \right\} + q^2
g_{\mu \nu}  \left\{ {1 \over 4} L + {1 \over 8} S \right\} \right] + \ldots
\nonumber \\
I_{\mu \nu \alpha} & = & \int {d^4 k \over (2 \pi)^4} {k_{\mu} k_{\nu}
k_{\alpha}
\over k^2 (k - q)^2 \left[ ( k - p')^2 + m^2 \right] } = \nonumber \\
& & {i \over 32 \pi^2 m^2} \left\{ \left[ p'_{\mu} p'_{\nu} p'_{\alpha}
\left\{ {1
\over
6} {q^2 \over m^2} L \right] \right\} \right.\nonumber \\
& &+ (p'_{\mu} p'_{\nu} q_{\alpha} + p'_{\mu} q_{\nu} p'_{\alpha} +
q_{\mu}
p'_{\nu} p'_{\alpha}) \left\{- {1 \over 3} {q^2 \over m^2} L - {q^2 \over
16m^2}
S \right\} \nonumber \\
& &+ (q_{\mu} q_{\nu} p'_{\alpha} + p'_{\mu} q_{\nu} q'_{\alpha} +
q_{\mu}
p'_{\nu} q_{\alpha} ) \left[ {1 \over 3} L \right] \nonumber \\
& &+ q_{\mu} q_{\nu} q_{\alpha} \left[ - L - {5 \over 16} S \right]
\nonumber \\
& &+ \left( g_{\mu \nu} p'_{\alpha} + g_{\mu \alpha} p'_{\nu} + g_{\nu
\alpha}
p'_{\mu} \right) \left[ - {q^2 \over 12} L \right] \nonumber \\
& &\left. + \left(g_{\mu \nu} q_{\alpha} + g_{\mu \alpha} q_{\nu} +
g_{\nu
\alpha}
q_{\mu} \right) \left[ {q^2 L \over 6} + {q^2 S \over 16} \right] \right\} +
\ldots
\end{eqnarray}

\noindent
In these latter set of integrals there are further nonanalytic terms with
higher
powers
of $q^2 / m^2$, which are not needed for the calculation of leading effects
in the
text.  Note that the nonanalytic parts of the integrals satisfy various "mass
shell"
constraints, such as

\begin{eqnarray}
q^{\mu} I_{\mu} & = & {q^2 \over 2} I + \ldots \nonumber \\
q^{\mu} I_{\mu \nu} & = & {q^2 \over 2} I_{\nu} + \ldots \nonumber \\
q^{\mu} I_{\mu \nu \alpha} & = & {q^2 \over 2} I_{\nu \alpha} + \ldots
\nonumber \\
q^{\mu} J_{\mu} & = & {q^2 \over 2} J + \ldots\nonumber \\
q^{\mu} J_{\mu \nu} & = & {q^2 \over 2} J_{\nu} + \ldots \nonumber \\
g^{\mu \nu} I_{\mu \nu} & = & 0 + \ldots \nonumber \\
g^{\mu \nu} I_{\mu \nu \alpha} & = & 0 + \ldots \nonumber \\
g^{\mu \nu} J_{\mu \nu} & = & 0 + \ldots \nonumber \\
\end{eqnarray}

\noindent where the ellipses denote analytic terms.  Here $p' \cdot q = q^2 /
2$ has
been used, following from $p^2 = (p' - q)^2$ on shell.  Some of these
relations
arise using $2k \cdot q = k^2 - (k - q)^2 + q^2$, and the fact that integrals
with
factors of $k^2$ or $(k - q)^2$ in the numerator can be shifted to remove all
dependence on $q$.

\begin{center}
{\bf Figure Captions}
\end{center}

\begin{description}
\item[Fig. 1.] a) A tree level vertex of order $\kappa^2 p^2$, b) a one loop
vertex of order $\kappa^4 p^4$.
\item[Fig. 2.] One graviton exchange for the Newtonian potential.  The
matter fields
are
represented by solid lines and wavy lines represent gravitons.
\item[Fig. 3.] The set of corrections included in the one particle reduceable
potential.

\item[Fig. 4.] The diagrams involved in the vertex correction.
\item[Fig. 5.] The diagrams in the graviton vacuum polarization.  The dotted
lines
indicate the
ghost fields.
\item[Fig. 6] Vertices required for the Feynman diagrams.
\end{description}

\end{document}